\documentclass[amssymb,amsmath,showpacs,aps,pre]{revtex4-1}
\usepackage{amsmath}
\usepackage{amsfonts}
\usepackage{amssymb}
\usepackage{dcolumn}
\usepackage{bm}
\usepackage[dvips]{graphicx}
\usepackage{epsfig}

\begin{document}

\title{Non-Markovian decay and dynamics of decoherence in private and public environments}
\author{A. D. Dente}
\author{P. R. Zangara}
\author{H. M. Pastawski}
\email{horacio@famaf.unc.edu.ar}
\affiliation{Instituto de F\'{i}sica Enrique Gaviola (CONICET) and Facultad
de Matem\'{a}tica, Astronom\'{i}a y F\'{i}sica, Universidad Nacional
de C\'{o}rdoba, Ciudad Universitaria, 5000, C\'{o}rdoba, Argentina}

\begin{abstract}
We study the decay process in an open system, emphasizing on the relevance of the
environment's spectral structure. Non-Markovian effects are included to
quantitatively analyze the degradation rate of the coherent evolution.
The way in which a two level system  is coupled to different environments is
specifically addressed: multiple connections to a single bath (public
environment) or single connections to multiple baths (private environments).
We numerically evaluate the decay rate of a local excitation by using the
Survival Probability and the Loschmidt Echo. These rates are compared to analytical results obtained from the standard Fermi Golden Rule (FGR) in Wide Band Approximation, and a Self-Consistent evaluation that accounts for the bath's memory in cases where an exact analytical solution is possible. We observe that the correlations appearing in a public bath introduce further deviations from the FGR as compared with a private bath.

\end{abstract}

\pacs{03.65.Yz, 03.65.Ta, 03.67.Pp}

\maketitle

\section{INTRODUCTION}

Quantum Information Processing (QIP) requires an efficient and precise control
on quantum dynamics \cite{divicenzo}. Decoherence conspires against this
objective as it leads to progressive and smooth destruction of the quantum
interferences \cite{zurek} within a\ characteristic time $\tau_{\phi}.$ The
main source of decoherence for solid-state spin systems such as quantum dots
\cite{vandersypen2,bluhm,marcus,imamoglu}, donors in silicon \cite{dzurak},
defects in diamond \cite{wrachtrup,awschalom,lukin2}, and solid state nuclear
magnetic resonance (NMR) \cite{VSB+01,cory1,alvarez1} is the uncontrolled spin
bath environment. While one can attempt different strategies such as quantum
error correction protocols \cite{error-correction-1,error-correction-2} and
dynamical decoupling \cite{d-decoupling}, their specific efficiency depends on
a deep understanding of how the environment behaves. Thus, understanding and
mitigation of decoherence is one of the current challenges for quantum science
and technology \cite{laflamme,degen,taylor}.

The interaction rate $1/\tau_{\mathcal{S-E}}$ between a quantum system
$\mathcal{S}$ and its environment $\mathcal{E}$ is typically evaluated from a
Fermi Golden Rule (FGR), which assumes that the environment has a Markovian
nature. Also, the interplay between the system time scales (e.g. $1/\omega
_{0}$) and that of the interaction ($\tau_{\mathcal{S-E}}$) could result in
striking effects. While weak interactions ($1/\tau_{\mathcal{S-E}}\ll
2\omega_{0}$) simply degrade dynamical interferences at a rate $1/\tau_{\phi
}\varpropto1/\tau_{\mathcal{S-E}}$, stronger ones may change the system's
response radically, leading to a quantum dynamical phase transition in its
dependence on $1/\tau_{\mathcal{S-E}}$ \cite{alvarez1}. Indeed, the
possibility of a non-analytic behavior, e.g. in the energy spectrum \cite{AndersonEXCHANGE}, appears because the system's
effective Hamiltonian is non-Hermitian \cite{rotter, hmp-physB}. This, in
turn, can be traced back to the fact that the environment has a number of
degrees of freedom $\mathtt{N}$ which can be considered infinite. As P.W.
Anderson put forth: \textquotedblleft more is different\textquotedblright%
,\ and new physical phenomena may appear when this thermodynamic limit
($\mathtt{N}\longrightarrow\infty$) is properly taken \cite{AndersonMORE}.
Indeed, there are several models of both non-Markovian and Markovian
environments \cite{leggett,danieli2,alv2007,hmp-physB} which show this dynamical phase transition as a function of the interaction strength.

\textbf{\ }While the Markovian approximation is sufficient for most
traditional applications, it leaves aside important memory effects and
interferences in the time domain. These result from the coherent interaction
between $\mathcal{S}$ and $\mathcal{E}$, and are becoming a topic of
increasing interest \cite{lukin}. The system-environment dynamics may go
through different temporal regimes: a quadratic decay at very short times
yields to the usual exponential FGR decay and much later it is transformed into an inverse power law decay (see Ref. \cite{elenacpl}). Here, we will
focus on the more relevant exponential regime, treated both with the FGR and with
a Self-Consistent (SC) FGR which accounts for bath memory effects. Once again, a deep understanding of
the environment's dynamics is central to identify the different regimes and to
foresee possible dynamical transitions.

A natural way to quantify the decoherence time $\tau_{\phi}$ is through the
degradation of specific interferences, e.g. Rabi oscillations
\cite{Muller-Kumar-Baumann-Ernst74,Danieli-SWAP-decoh} or mesoscopic echoes
\cite{madi-ernst1997,alv2010,mesoECO-theory,mesoECO-exp}. Alternatively, the
implementation of a time reversal procedure, the Loschmidt Echo (LE)
\cite{jalabert-hmp}, allows the evaluation of the decoherence time by
measuring the reversibility of the system's dynamics in presence of an
uncontrolled environment. The LE can be accessed experimentally in spin
systems \cite{usaj2,patricia98}, confined atoms \cite{davidson}, microwave
excitations \cite{gorinMicroWave}, etc., and has become a powerful tool for
quantifying decoherence, stability and complexity in dynamical processes in
several physical situations \cite{prosen,Jacquod}.

In this article we have the aim to quantify the role of bath's memory as well as specific correlations in the  $\mathcal{S-E}$ interaction on decoherence. With this purpose, we consider the evolution of two coupled spins in the
$\uparrow\downarrow$ and $\downarrow\uparrow$ configurations in the presence of different spin
environments with a fully characterized coherent dynamics \cite{axel}. The system's Rabi
oscillations \cite{Cohen-Tannoudji} can act as a SWAP gate, and after
appropriate mappings, this boils down to an excitation that jumps between two
degenerate states $A$ and $B$ with a rate determined by the coupling ``constant'' $V_{AB}$, which can be
switched at will. Starting on state $A$, the return probability oscillates
with $\omega_{0}=2V_{AB}/\hbar$, the Rabi frequency. In
order to understand the incidence of environmental memory effects in the rate
$1/\tau_{\phi}$, we must consider the relation
between the system's time scale (typically ruled by $\hbar/V_{AB}$) and the
bath's inner excitation spreading time scale as determined by the density of
directly connected states $\hbar N_{1}$, which, in general, is just a Local Density of States (LDoS) \cite{Elena2D}.
Also important is the specific form in which the
system couples to the bath: each site may be coupled to a different environment
(private bath) or both sites could share the same environment (public bath), and this would allow different correlations.
A quantitative comparison between rates in these cases will enable a  qualitative
interpretation in terms of the bath's spectral structure and the effects of public and private $\mathcal{S-E}$ interactions. This should deepen
our understanding about how the involved correlations modify the degradation
rate $1/\tau_{\phi}$. Indeed, if the system's and the bath's time scales are
similar, the problem cannot be treated within the Markovian paradigm of a
\textquotedblleft slow variable interacting with a fast equilibrating
background\textquotedblright. It requires to be carefully addressed beyond the
FGR.  The appearance of mechanisms of correlated coupling in public environments has been previously pointed in the literature
of open quantum systems, particularly in terms of the spin-boson models
\cite{ekert, breuer2007}. It was shown \cite{breuer2007}, at least for a
simple model of a $\mathtt{N}-$qubit register, that the decoherence increases linearly
with $\mathtt{N}$ for independent reservoirs, while it grows with the square
$\mathtt{N}^{2}$ for a collective environment. This suggests that such a public
interaction may lead to a strong amplification of decoherence. On the other
hand, it has been pointed in the literature of error correction protocols that
symmetric (public) $\mathcal{S-E}$ coupling can be exploited to design states
that are hardly corrupted by such a coherent environmental noise
\cite{zanardi-noise}. Also, in recent years the role
of a common environment in the entanglement correlations within a system
\cite{buchleitner,paz-roncaglia}, and the possibility of creating and
manipulating those correlations by environment-mediated interactions
\cite{davidovich} have been explored.

It is important to notice that spectral correlations within the bath, as described above, may become
quite cumbersome to describe in statistical terms. However,
realistic Hamiltonian models of the bath, as proposed here, should allow a simple and natural
description of such correlations.
Thus, the cases we analyze here can be casted directly into 1-D spin systems interacting by
means of a planar ($XY$ or flip-flop) or as chains of spins interacting with a double-quantum (flip-flip /
flop-flop) effective Hamiltonians.
Indeed, the spin-fermion mapping provided
by the Jordan-Wigner Transformation (JWT) \cite{jwt} has been successfully
exploited to predict spin polarization dynamics in linear chains and rings
\cite{mesoECO-theory} and results in full agreement with experiments
\cite{mesoECO-exp,madi-ernst1997}.
Furthermore, it allowed addressing polarization
spin dynamics in homogeneous chains \cite{danieli1,Danieli-SWAP-decoh,elenacpl} and dynamics of multiple
spin coherences  \cite{feldman1,feldman2,cory2,elenaMQC} by mapping
them to fermionic excitations. In both cases, the local excitation is
identified with a single fermion propagating in a tight-binding linear chain,
and one can assume that the environment is described by an identical chain.
Additionally, the dynamics of  a tight-binding model can be mapped, sometimes quite straightforwardly, to describe
the propagation of excitations in several scenarios like classical and
quantum coupled oscillators \cite{Economou}, plasmonic wave guides
\cite{plasmonics-Bustos-Marun}, sound propagation
\cite{Calvo-vibrations-electrons}, spin-boson interactions and other models
used for decoherence in quantum information \cite{Paz-Zhang, Bartels}.

A fundamental issue about integrability of the chosen models is that they enable
 analytical expressions valid in the thermodynamic limit of an infinite
number of spins.
In fact, an analytically tractable bath means the capability to sum up an
otherwise divergent perturbation series into a complex self-energy through the
Dyson Equation, which only then is considered in the FGR approximation. Additionally, having
a smooth LDoS, which arises from a continuous spectrum ($\mathtt{N}%
\longrightarrow\infty$), avoids spurious resonances that might appear in
finite systems. Besides, we will see that a well defined curvature of the LDoS will
be central to quantify the corrections to the FGR.

This paper is organized as follows. In Sec. \ref{sec:TB-model} we present the
$\mathcal{S-E}$ tight-binding models that yields a fully solvable quantum
dynamics (the underlying spin-fermion mapping is summarized in the Appendix
\ref{Apendice-1}). In Sec. \ref{sec:Numerical-Analytical-Tools}, the notions
of Survival Probability and local Loschmidt Echo are presented on the face of
their numerical implementation. We deal analytically with every case by working with the exact Green's Function in each case. Further details related to the GF's poles are discussed in Appendix \ref{Apendice-2}. In Sec. \ref{sec:Results},
the analytical decay rates obtained with the FGR and the SC-FGR are presented
for each case under consideration. These rates are compared with those
computed numerically from the dynamics, contrasting the results of the Survival
Probability (SP) and the Loschmidt Echo (LE) of the local excitation. All
these magnitudes are analyzed under the light of the knowledge of the
environment memory effects as provided by its spectral structure (LDoS). In
the last section, further discussions and conclusions are presented to argue
how the public bath is more effective to distort the decay process away from
the usual Fermi Golden Rule (energy independent rate).

\section{TIGHT BINDING MODEL FOR EXCITATION DYNAMICS}

\label{sec:TB-model}

\begin{figure}
    \centering
     \includegraphics[width=0.8\textwidth]{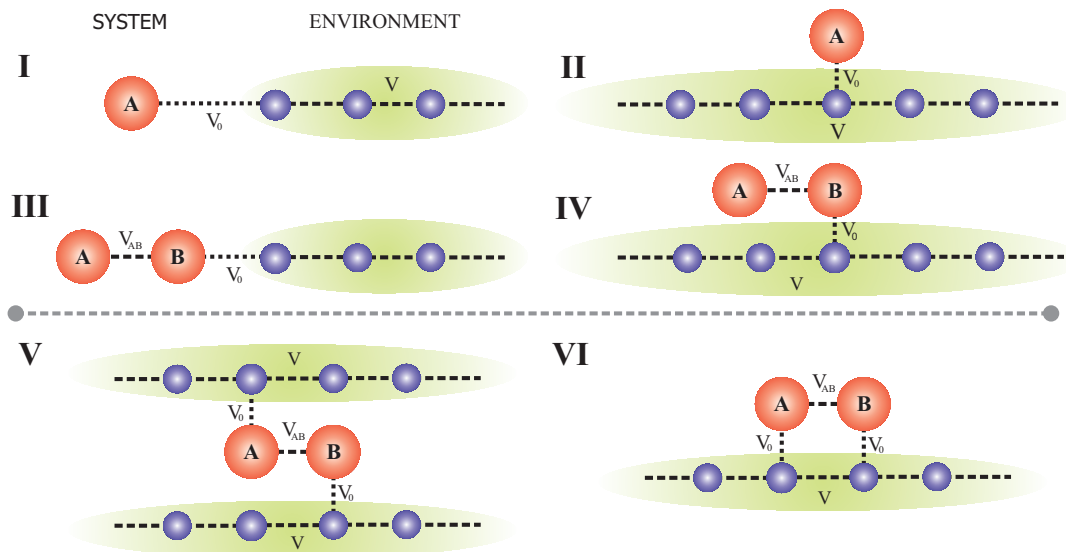}\\
      \caption{(Color Online) Non interacting fermion modelization.
      \textbf{I}) A single site state connected to a semi
infinite chain. \textbf{II) }One site coupled to an infinite chain. Since $%
V_{0}\ll V$, both cases can be treated by a simple decay process, given by
the FGR. \textbf{III}) Two sites; the initial configuration is given by a
particle in the site \textbf{A}. The sites are coupled to a semi infinite
chain by only one of them. \textbf{IV)} Same as \textbf{III}), but with an
infinite linear chain standing for the bath.\textbf{V) }Two sites coupled
both to different infinite chains (private environments). \textbf{VI)} Two
sites coupled to the same infinite chain (public environment).}%
    \label{Fig Esquema de Sistemas}%
\end{figure}

As pointed above, while our motivation lies mainly on spin dynamics under
flip-flop interactions, we use the spin-fermion mapping to cast it in terms of
tight-binding models that apply to a wide variety of systems. Thus, we leave
to Appendix \ref{Apendice-1} a brief outline of how this mapping is achieved.
Here, we present the models and analyze them in terms of straightforward
single particle physics. These are basically variations of tight-binding
infinite linear chains where the bath's memory can be fully characterized. Of
course, the cases where the interactions network topology has branching points
or loops would preclude the simple back transformation into spin systems.
However, even in these situations some of the physics of the memory would
remain. The general situations are sketched in
Fig.\ref{Fig Esquema de Sistemas}.%

For the cases considered in this article, the whole Hamiltonian can be written as:%

\begin{equation}
\hat{H}=\hat{H}_{\mathcal{S}}+\sum_{\nu}\hat{H}_{\mathcal{E}_{\nu}}+\hat{H}_{\mathcal{S-E}%
},\label{h1}%
\end{equation}
where:%

\begin{equation}
\hat{H}_{\mathcal{S}}=E_{A}\hat{c}_{A}^{\dag}\hat{c}_{A}^{{}}+E_{B}\hat{c}%
_{B}^{\dag}\hat{c}_{B}^{{}}-V_{AB}(\hat{c}_{A}^{\dag}\hat{c}_{B}^{{}}+\hat
{c}_{B}^{\dag}\hat{c}_{A}^{{}}),\label{h2}%
\end{equation}
with $\hat{c}_{s}^{\dag}$ and $\hat{c}_{s}$ ($s\in\left\{  A,B\right\}  $),
the creation and destruction operators for fermions and $V_{AB}$ is the
hopping amplitude that defines the system's only dynamical time scale provided
that $E_{A}=E_{B}$. Each bath $\mathcal{E}_{\nu}$ ($\nu$=1,2), with a spectral
bandwidth of $4V,$ is described by the Hamiltonian:%

\begin{equation}
\hat{H}_{\mathcal{E}_{\nu}}=\sum_{n=n_{_{\nu}}}^{\infty}E_{\nu n}\hat{c}_{\nu
n}^{\dag}\hat{c}_{\nu n}^{{}}-V(\hat{c}_{\nu,n+1}^{\dag}\hat{c}_{\nu,n}^{{}%
}+\hat{c}_{\nu n}^{\dag}\hat{c}_{\nu.n+1}^{{}}).\label{h3}%
\end{equation}
Choosing the site energies with identical values: $E_{\nu,n}=0$ ensures a
continuum spectrum, while the election $E_{A}=E_{B}=0$ will further simplify
the analysis. Two different alternatives for the spectrum dynamics arise when
$n_{\nu}=$ $1$ (semi-infinite linear chain) and\textbf{\ }$n_{\nu}=-\infty$
(infinite linear chain). The system-bath Hamiltonian $\hat{H}_{\mathcal{S-E}}$
depends on how we couple our two-site system to the environment (linear
chain), but in general it will be of the form:%

\begin{equation}
\hat{H}_{\mathcal{S-E}}=-V_{0}\left(  \hat{c}_{A}^{\dag}\hat{c}_{\nu i}^{{}%
}+\hat{c}_{\nu i}^{\dag}\hat{c}_{A}^{{}}+\hat{c}_{B}^{\dag}\widehat{c}_{\mu
j}^{{}}+\hat{c}_{\mu j}^{\dag}\hat{c}_{B}^{{}}\right)  ,\label{h4}%
\end{equation}
where $\nu i$ and $\mu j$ label sites $i$ and $j$ in the environments $\nu$
and $\mu$ respectively.

It is crucial to stress that $V_{AB},$\ $V_{0}$\ and $V$\ determine the
relevant time scales of the whole problem. The first two give the rate of
hopping from site $A$ to site $B$ and to the environment, respectively. The
third is the jumping rate between sites in the environment. An
\textquotedblleft irreversible\textquotedblright\ decay to the environment,
and hence the Fermi Golden Rule, implies that the unperturbed isolated system
state has zero overlap with eigenstates of $\mathcal{S}+\mathcal{E}.$\ For
this perturbation theory break down, the interaction with each environment
eigenstate, $V_{0}/\sqrt{\text{\texttt{N}}}$ must be much greater than the
spacing between adjacent levels, of about $V/\mathtt{N},$ i.e. the interaction
time scale $\hbar/V_{0}$ must be lower than the environment's Heisenberg
time $\hbar\mathtt{N}/V$. Imposing $V_{0}\ll V_{AB}$ for all the cases we treated (weak coupling regime), we ensure the smooth degradation of the system's coherent evolution.
When $V\approx V_{AB}$ the memory effects
characterizing a \emph{non-Markovian} situation lead to a very rich dynamical
behavior. In the opposite limit, if $V\gg V_{AB}$ then the validity of the FGR
(\emph{Markovian} situation) is expected to be recovered. Typically, this last
situation will be represented in this work by a hopping $V=5V_{AB}$.

\section{NUMERICAL AND ANALYTICAL TOOLS}

\label{sec:Numerical-Analytical-Tools}

\subsection{SURVIVAL PROBABILITY AND LOSCHMIDT ECHO}

\label{subsec:NumerialSol}

Two kind of measures for dynamical degradation are employed in this work: the
Survival Probability (SP) $P_{AA}(t)$ and the local Loschmidt Echo. The SP is
defined as%

\begin{equation}
P_{A,A}(t)=\left\vert \left\langle A\right\vert \exp\left[  -\mathrm{i}\hat
{H}t/\hbar\right]  \left\vert A\right\rangle \Theta(t)\right\vert
^{2},\label{s7}%
\end{equation}
where $\Theta(t)$ is the Heaviside step function, and the Hamiltonian $\hat
{H}$ is defined by Eq. \ref{h1}. It measures the probability of finding a
particle in site $A$ at time $t$, provided that the system has had a particle
in the same site at time $t=0$. In spin systems, this is a spin
autocorrelation function (see Eq. \ref{s1} and \ref{s6} in Appendix
\ref{Apendice-1}). The whole evolution of the system as reflected in the SP,
is affected by a decay process, which is not trivial to separate from the
intrinsic dynamics. Thus, to quantify decoherence, one relies on the
observation of specific features as natural recurrences (Rabi oscillations or
mesoscopic echoes) that appear at specific times. This limits the used time
windows and limits the detailed assessment of $\mathcal{S-E}$ dynamics.
Although it is not the perfect tool to quantify the effects of the
environment, the SP behaves as a probe that reflects the overall dynamical process.

With the purpose of get a continuous access to the $\mathcal{S-E}$ dynamics
that better reflects the environmental memory effects, we focus our attention
on the LE. This measure has been used in the last years in different
physical scenarios (both experimental
\cite{patricia98,usaj2,gorinMicroWave,davidson} and theoretical
\cite{zurek2,prosen,jalabert-hmp}) in order to explain the behavior of the
decoherence characteristic time. In general, the LE provides a direct measure
of the decoherence process due to the environment, and even though it depends on the
nature of the system's intrinsic dynamics, it does not depend much on its
details. Its usual dynamical behavior presents an exponential decay regime
\cite{comment1}, which will be used to characterize the destruction of the
system's coherent dynamics.

The LE relies on the time reversal of the system's evolution and, in the
present scenario, it has a direct physical interpretation. The LE can be
understood as the measure of the amount of polarization returned
to\textit{\ local }site where it started. The controlled quantum dynamics is
separated in two stages. First, the initial local excitation (particle in site
$A$) evolves during a time $t_{1}$, and then a time reversal procedure is
applied during a period time ($T-t_{1}$) which reverses the system's dynamics
($\hat{H}_{\mathcal{S}}\rightarrow-\hat{H}_{\mathcal{S}}$). It is important to
note that the bath's dynamics and the $\mathcal{S-E}$ interactions ($\hat
{H}_{\mathcal{E}}$ and $\hat{H}_{\mathcal{S-E}}$ respectively) remain
unreversed during the backward evolution. This partial control results in a
non-reversed perturbation $\hat{\Sigma}=(\hat{H}_{\mathcal{E}}+\hat
{H}_{\mathcal{S-E}})$ acting in both periods. Finally, the probability of
finding the particle in site $A$ forms a Loschmidt echo provided that
$t_{1}=T/2$:%

\begin{equation}
M_{LE}(T)=\left\vert \left\langle A\right\vert \exp\left[  -\mathrm{i}%
(-\hat{H}_{\mathcal{S}}+\hat{\Sigma})(T-t_{1})/\hbar\right]  \exp\left[
-\mathrm{i}(\hat{H}_{\mathcal{S}}+\hat{\Sigma})t_{1}/\hbar\right]  \left\vert
A\right\rangle \right\vert ^{2}.\label{s8}%
\end{equation}
Clearly, in the case where the system is isolated, the local LE will have a
steady value of $1$. This means that the system is fully reversible. On the
other hand, if the system is coupled to the infinite and continuous environment's
spectrum, both the forward and backward effective Hamiltonians, $\hat
{H}_{\mathcal{S}}+\hat{\Sigma}$ and $-\hat{H}_{\mathcal{S}}+\hat{\Sigma}$
respectively, become non-Hermitian. Thus, the LE decays with a characteristic
rate, i.e., our reversal procedure has not been able to recover the excitation
spread to the environment. The net decay of the norm of the state simply means
the decay of the coherent part and then describes the non-trivial part of the
Loschmidt Echo.

\begin{figure}
    \centering
      \includegraphics[width=0.8\textwidth]{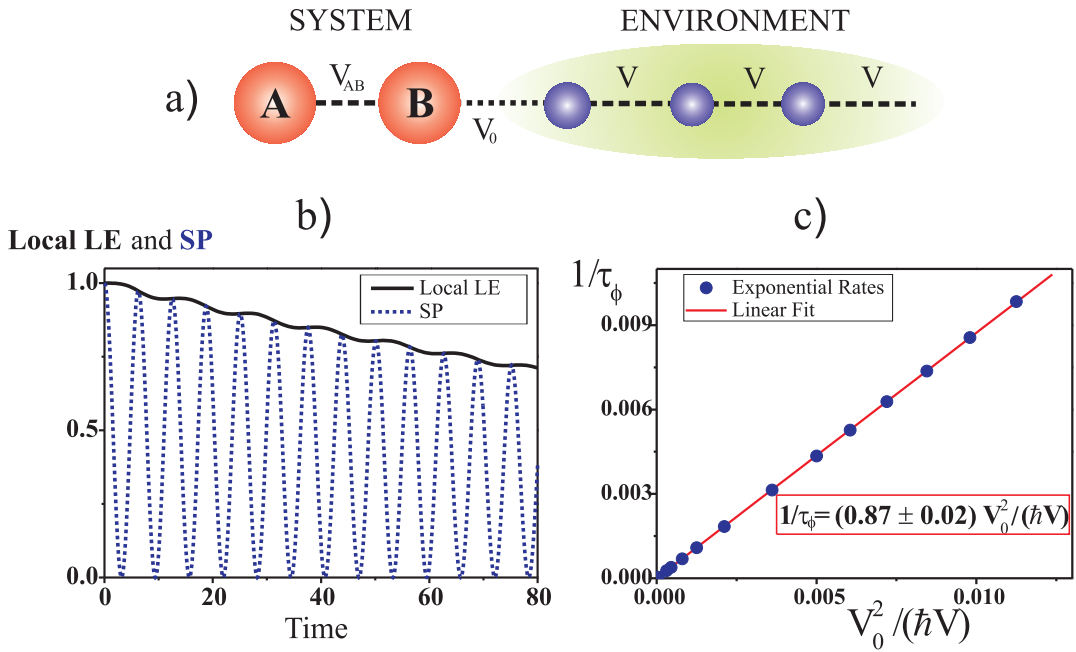}\\
      \caption{(Color Online) \textbf{a})
Typical tight binding model considered. \textbf{b}) SP (dotted line) and the
local LE (solid line) plots vs. time (in units of $\hbar/V_{AB}$). \textbf{c}) Characteristic decay rate
(exponential decay regime) as a function of $V_{0}^{2}/\hbar V$ (both in units of $V_{AB}/\hbar$). The points were
obtained by the variation of the coupling constant $V_{0}$.}%
\label{Fig SP and LLM}%
\end{figure}

For both survival probability and Loschmidt Echo evaluation, we diagonalize
the Hamiltonian to obtain the evolution for every time. We use sufficiently
large chains to approximate the nature of infinite ones. Indeed, the evolution
times considered in this work are short enough to ensure the absence of
dynamical finite-size effects (e.g. mesoscopic echoes appearing at the
environment's Heisenberg time). A typical system is presented
in Fig. \ref{Fig SP and LLM}-\textbf{a} (the same model was deeply analyzed in
Ref. \cite{axel}).

We are interested in studying the decay rates as a function of the
$\mathcal{S-E}$ coupling parameter
$V_{0}$ (from Hamiltonian of Eq. \ref{h4}). We fit the
dynamics of the local LE  with exponential decay function, which for the case plotted in Fig.
\ref{Fig SP and LLM}-\textbf{b}, turns out to be the envelope of the Rabi
oscillations in the SP and has the advantage of having a monotonous behavior. However, one should be aware that for an arbitrary system, the SP envelope does not
necessarily match the local LE (the characteristic rates can be different). Every rate we obtain, is related to a particular choice of $V_{0}$, and we want to use them to asses the conditions of validity of a FGR regime:%

\begin{equation}
\frac{1}{\tau_{\phi}}\simeq\frac{2\pi}{\hbar}\left(  \hat{H}_{\mathcal{S-E}%
}\right)  ^{2}N_{1},\label{s9}%
\end{equation}
where $\left(  \hat{H}_{\mathcal{S-E}}\right)  ^{2}$ is a characteristic
second moment of the $\mathcal{S-E}$ interaction and $N_{1}$ represents an
appropriate density of directly connected states. Thus, we plot the decay rates as a function of $V_{0}^{2}/\hbar V$
(Fig. \ref{Fig SP and LLM}-\textbf{c}). It is interesting to note
that because of the linear chain topology this second moment coincides with
$V_{0}^{2}$ and $N_{1}$ can be identified with a local density of state at the
first site of the chain. In a general environment this correspondence can be
assigned through a Lanczos transformation \cite{Elena2D}.%

\subsection{ENVIRONMENT'S EFFECT IN A SELF-CONSISTENT FERMI GOLDEN RULE}
\label{subsec:Analytical-Approach}

\begin{figure}
    \centering
      \includegraphics[width=0.4\textwidth]{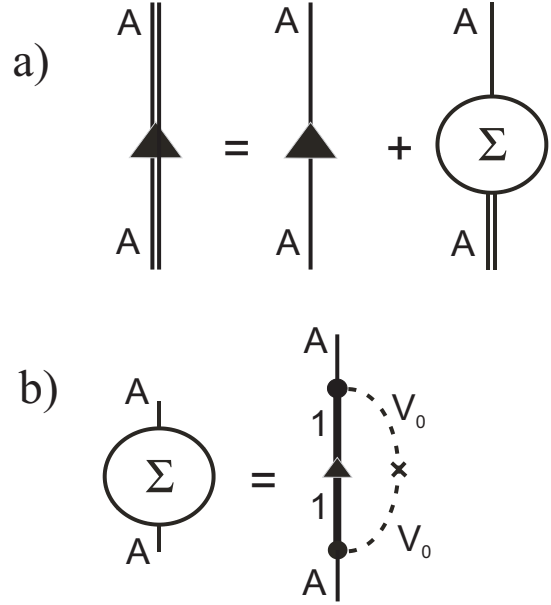}\\
      \caption{\textbf{a}) Diagrammatic
representation for the retarded GF at site $A$, in the form of a Dyson's
equation. The interaction with the environment is to infinite order in the
self-energy given in (\textbf{b}) (see Eq. \protect\ref{s14}). Simple lines
with arrows are exact GF in absence of $\mathcal{S-E}$ interactions. \textbf{%
b}) Self-energy diagram sums all orders in the hopping to the environment.
Thick line with arrows is the exact propagator at a point of the isolated
bath, denoted by $\overline{G}_{11}^{R}$ (see text).}%
\label{feynman diag}%
\end{figure}

For the study of the analytical solutions we use the Green's Function (GF)
formalism. In this framework it is possible to obtain the overall dynamics of
the system by looking at the behavior of the GF poles \cite{elenacpl,axel}.
Any element of the retarded GF can be obtained from the Fourier transform of the full
propagator:

\begin{align}
G_{AA}^{R}(\varepsilon) &  =\lim_{\eta\rightarrow0}{\displaystyle\int
\limits_{-\infty}^{\infty}}\left\langle A\right\vert \exp\left[
-\mathrm{i}~\left(  \hat{H}-\mathrm{i}\eta\hat{I}\right)  ~t/\hbar\right]
\left\vert A\right\rangle \Theta(t)\exp\left[  +\mathrm{i}\varepsilon
t/\hbar\right]  \mathrm{d}t\nonumber\\
&  =\frac{1}{\varepsilon-E_{A}-\Sigma(\varepsilon)}=\frac{1}{\varepsilon
-E_{A}-\Delta(\varepsilon)+\mathrm{i}\Gamma(\varepsilon)},\label{s14}%
\end{align}
where $\Sigma(\varepsilon)$ is the appropriate self-energy with real and
imaginary parts $\Delta(\varepsilon)$ and $-\Gamma(\varepsilon)$ respectively.
The self-energy operator is diagrammatically presented in Fig.
\ref{feynman diag}. The bath memory is contained in their dependence on
$\varepsilon,$ which arises from the bath's exact Green's Function $\overline{G}_{11}%
^{R}(\varepsilon)$ at the directly connected site. Indeed, $\Sigma\left(
\varepsilon\right)  $ is given by $\left\vert V_{0}\right\vert ^{2}%
\overline{G}_{11}^{R}\left(  \varepsilon\right)  $, see Fig.
\ref{feynman diag}-\textbf{b}. All the cases considered in this work (see Fig.
\ref{Fig Esquema de Sistemas}) can be reduced to the self-energy of a semi
infinite linear chain ($n_{\nu}=$ $1$), i.e.%

\begin{equation}
\Sigma(\varepsilon)=\dfrac{V_{0}^{2}}{V^{2}}\left[  \dfrac{\varepsilon}%
{2}\text{ }-\mathrm{i}\sqrt{V^{2}-\left(  \frac{\varepsilon}{2}\right)  ^{2}%
}\right]  \text{\ \ for }\left\vert \varepsilon\right\vert \leq2\left\vert
V\right\vert .\label{delta}%
\end{equation}
For the infinite linear case ($n_{\nu}=-\infty$), the previous expressions
must be multiplied by $2$.

Notice that $\Sigma(\varepsilon)$ plays the role of the influence functional
in the Feynman path integral formulation usually used to deal with memory
effects for bosonic baths\cite{Ingold-tutorial}. However, in such cases there
are many more free parameters than in our case, i.e. the bath spectral
density, the coupling strength with each mode and the temperature that fixes
the bath occupation. In our spin environment model however, the last is
simplified by the high temperature limit, while the two first become naturally
determined by specific sum rules arising from the physical Hamiltonian we select.%

We remark that the complex GF poles are consequence of the unbounded nature of
the system which prevents mesoscopic echoes \cite{mesoECO-theory} and
Poincare's recurrences. Also it is important to point out that we are working
with a Hamiltonian in a parametric region holding a continuous spectrum where
no localized modes appear.

To get a better understanding of the connection between the dynamical behavior
and the poles of the GF, we span the initial condition in the Survival
Probability (see Eq. \ref{s7}) in terms of the energy eigenstates:%

\begin{align}
P_{AA}(t) &  =\left\vert \Theta(t)\sum_{k=1}^{\infty}\left\vert \left\langle
\psi_{k}|A\right\rangle \right\vert ^{2}\exp\left[  -\mathrm{i}\varepsilon
_{k}t/\hbar\right]  \right\vert ^{2}\nonumber\\
&  =\left\vert \Theta(t){\displaystyle\int\limits_{-\infty}^{\infty}%
}\mathrm{d}\varepsilon\left[  \sum_{k=1}^{\infty}\left\vert \left\langle
\psi_{k}|A\right\rangle \right\vert ^{2}\delta(\varepsilon-\varepsilon
_{k})\right]  \exp\left[  -\mathrm{i}\varepsilon_{k}t/\hbar\right]
\right\vert ^{2}.\label{s11}%
\end{align}
The term in brackets is identified as the Local Density of States (LDoS)
$N_{A}(\varepsilon)$ at the site $A$:%

\begin{align}
N_{A}(\varepsilon) &  =-\frac{1}{\pi}\operatorname{Im}{\displaystyle\int
\limits_{-\infty}^{\infty}}\mathrm{d}tG_{AA}^{R}(t)\exp\left[  -\mathrm{i}%
\varepsilon t/\hbar\right] \nonumber\\
&  =-\frac{1}{\pi}\operatorname{Im}G_{AA}^{R}(\varepsilon).\label{s12}%
\end{align}
Therefore, we can identify Eq. \ref{s7} as the LDoS Fourier transform:%

\begin{equation}
P_{A,A}(t)=\left\vert \Theta(t){\displaystyle\int\limits_{-\infty}^{\infty}%
}\frac{\mathrm{d}\varepsilon}{2\pi\hbar}N_{A}(\varepsilon)\exp\left[
-\mathrm{i}\varepsilon t/\hbar\right]  \right\vert ^{2}.\label{s13}%
\end{equation}
The last expression can be numerically and analytically computed once we know
the GF in the energy representation. As a matter of fact, since we have
$\Sigma(\varepsilon)$, and hence $N_{A}(\varepsilon),$ explicitly, we can
compute the dynamics from the Fourier Transform mentioned. But, we prefered to
obtain the numerical evolution (by exact diagonalization, as
explained in previous section) of a finite environment. The interest in this
last method arises from the fact that its use can be directly generalized to
more complex systems\ (many-body), where an exact analytical solution is not
accessible. The analytical alternative based on the evaluation of the GF poles
will enable to summarize the decay rate in a simple expression, as it is shown below.

The characteristic decay of $P_{A,A}(t)$ is determined by the bath's LDoS,
$N_{1}\left(  \varepsilon\right)  $. This last can be obtained from the
isolated bath's exact GF at site $1$ (i.e. $\overline{G}_{11}^{R}$),%

\begin{equation}
N_{1}\left(  \varepsilon\right)  =-\frac{1}{\pi}\operatorname{Im}\overline
{G}_{11}^{R}\left(  \varepsilon\right)  .\label{gf8}%
\end{equation}
Hence, we recall that $N_{1}\left(  \varepsilon\right)  $ plays the most
relevant role in $\mathcal{S-E}$ dynamics as was quoted in Ref. \cite{axel}.
In some systems the LDoS can be factorized as $N_{A}(\varepsilon
)=N_{1}(\varepsilon)\times L_{1}(\varepsilon)\times L_{2}(\varepsilon)$, where
$L_{i}(\varepsilon)$ are Lorentzian functions\ (LFs) related to the real and
imaginary parts (denoted by $\Delta_{0}$ and $\Gamma_{0}$ respectively) of the
GF poles. Hence, the convolution theorem applied to Eq. \ref{s13} leads to a
characteristic decay of $P_{A,A}(t)$ ruled by $\Gamma_{0}$. In fact, the decay
parameter of the exponential regime is given by $1/\tau=2\Gamma_{0}/\hbar$.
This is what we call self-consistent Fermi Golden Rule (SC-FGR)
\cite{elenacpl}.

With the purpose of finding $\Gamma_{0}$, we focus on the Hamiltonian
\ref{h1}, and follow the continued-fraction procedure described in Ref.
\cite{Pasta-Medina}. For each type of $\mathcal{S-E}$ coupling (see Fig.
\ref{Fig Esquema de Sistemas}), it is necessary to recalculate the poles of
the GF. Due to the small number of poles of the systems, it was feasible to
obtain the analytical solutions for all the cases presented in this work.
Also, we address the behavior of $\Gamma_{0}$ as a function of $V_{0}\ll1$.
Indeed, we proceed to expand the solution near $V_{0}\simeq0$.

It is important to notice that in the Taylor expansion, the linear and zero
order terms vanish. Therefore, the imaginary part has $V_{0}^{2}$ as the first
non trivial term. As expected for a FGR, this is in strong agreement with Eq.
\ref{s9}, where we identify the order $V_{0}^{2}$ as the second moment of the
$\mathcal{S-E}$ interaction (in general denoted by $\left\Vert \hat
{H}_{\mathcal{S-E}}\right\Vert ^{2}$). In the next section, we present the
corresponding values for $1/\tau=2\Gamma_{0}/\hbar$, expressed in the first
non trivial order\ (as we said, the $2^{nd}$), for each case considered.

The usual alternative (easier and cheaper) to the presented scheme (SC-FGR) is
the simple FGR, which is equivalent to evaluate the Green's function in a
first pole approximation:
\begin{align}
\left[  G_{AA}^{R}\left(  \varepsilon\right)  \right]  ^{-1} &  \simeq
\varepsilon-E_{A}-\Sigma(E_{A})\label{s15}\\
&  =\varepsilon-E_{A}-\Delta(E_{A})+\mathrm{i}\Gamma(E_{A})\label{s16}%
\end{align}
Since this yields an $\varepsilon$-independent rate, it can be understood as a
Wide Band Approximation (WBA) \cite{PascazioFGRbook,Pascazio-pa99,Pascazio-pla98}. This approximation
would imply neglecting any signature of dynamics and memory effects in the
environment. Also, it misses some striking dynamical behaviors appearing at
long times as the survival collapse \cite{elenacpl,axel} and the subsequent
power law decay \cite{Khalfin,Ghirardi,GarciaCaderon-Moshinsky}.

As a matter of fact, the WBA is represented by the condition of $V\gg V_{AB}$.
Under this assumption, the environment acquires fast dynamics and the system
does not receive any return from it (Markovian limit). In general, the common
FGR has the form expressed in Eq. \ref{s9}. There, the last factor ($N_{1}$)
stands for the LDoS of the directly connected states, and by the application
of the WBA, it is evaluated in the middle of the band spectrum (in our case,
$\varepsilon=0$).

In this work it is important to remember that the LDoS varies from the
semi-infinite chain ($n_{\nu}=$ $1$, surface state) to the infinite chain
($n_{\nu}=-\infty$, bulk state) as follows,%

\begin{align}
N_{1s}\left(  \varepsilon\right)   &  =\frac{1}{\pi V^{2}}\left(  V^{2}%
-\frac{\varepsilon^{2}}{4}\right)  ^{1/2}\Theta\left[  2V-|\varepsilon| \right],\label{ldos-s}\\
N_{1b}\left(  \varepsilon\right)   &  =\frac{1}{2\pi}\left(  V^{2}%
-\frac{\varepsilon^{2}}{4}\right)  ^{-1/2}\Theta\left[ 2V-|\varepsilon| \right].\label{ldos-b}%
\end{align}
From the previous spectral structures, we stress the presence of van Hove
singularities, which play an important role for long time dynamical behavior
\cite{Khalfin,elenacpl,axel}. Indeed, the critical exponent characterizing the
van Hove singularity can be related to the dimensionality of the space where
the quantum excitation diffuses. Also, the two different convexities in the
LDoS, near the middle of the band spectrum, will be of great relevance in the
following discussions.

\section{DECAY RATES: FGR AND BEYOND}

\label{sec:Results}

In this section we expose the main results of our work, summarized in Table
\ref{tabla-1}. The decay rates are presented as function of $V_{0}^{2}/\hbar
V$, for every case analyzed (Fig. \ref{Fig Esquema de Sistemas}) and each
approach employed (SP degradation, local LE degradation, WBA-FGR, and SC-FGR).

\begin{table}[ptb]
\caption{Degradation rates for all the methods analyzed (Survival
Probability and local Loschmidt Echo degradation, Wide Band Approximation, and
Self Consistent FGR). The cases are described in Fig.
\ref{Fig Esquema de Sistemas}. If the system's and bath's time scales are equal,
the physical situation is strickly non-Markovian. We point to restore
Markovianity in cases where both scales differ by a factor of five.}%
\label{tabla-1}%
\centering
\resizebox{1.0\textwidth}{!}{
\begin{tabular}{|c|c|c|c|c|}
\hline
System & SP degradation rate $\left[\frac{V_{0}^{2}}{\hbar V}\right]  $ & Local LE degradation rate $\left[  \frac{V_{0}^{2}}{\hbar V}\right]  $ & WBA $\left[  \frac{V_{0}^{2}}{\hbar V}\right]  $ & SC-FGR $\left[  \frac{V_{0}^{2}}{\hbar V}\right]  $\\
\hline
$\mathbf{I}$ & $2.04\pm0.05$ & $2.04\pm0.05$ & $2$ & $2$\\
\hline
$\mathbf{II}$ & $1.00\pm0.02$ & $1.00\pm0.02$ & $1$ & $1$\\
\hline
$\mathbf{III}$ - ($V=V_{AB}$) & $0.88\pm0.05$ & $0.88\pm0.05$ & $1$ & $0.87$\\
\hline
$\mathbf{III}$ - ($V=5V_{AB}$) & $1.00\pm0.02$ & $1.00\pm0.02$ & $1$ &$0.995$\\
\hline
$\mathbf{IV}$ - ($V=V_{AB}$) & $0.56\pm0.02$ & $0.56\pm0.02$ & $0.5$ &$0.577$\\
\hline
$\mathbf{IV}$ - ($V=5V_{AB}$) & $0.50\pm0.02$ & $0.50\pm0.02$ & $0.5$ &$0.502$\\
\hline
$\mathbf{V}$ & $1.16\pm0.03$ & $1.16\pm0.03$ & $1$ & $1.15$\\
\hline
$\mathbf{VI}$ - ($V=V_{AB}$) & $1.71\pm0.04$ & $1.20\pm0.04$ & $1$ & $1.732 $(\textit{forward}) and $0.577$ (\textit{backward})\\
\hline
$\mathbf{VI}$ - ($V=5V_{AB}$) & $1.11\pm0.03$ & $1.02\pm0.03$ & $1$ & $1.106$(\textit{forward}) and $0.904$ (\textit{backward})\\
\hline
\end{tabular}
}\end{table}

The rates predicted in the WBA column of Table \ref{tabla-1}, correspond to
the direct evaluation of Eq. \ref{s9}, i.e.:%

\begin{equation}
\frac{1}{\tau_{\phi}}\simeq\frac{2\pi}{\hbar}V_{0}^{2}N_{1\lambda}%
(\varepsilon=0),\label{s10}%
\end{equation}

where $\lambda$ stands for $s$ (surface, for semi infinite chains) or $b$
(bulk, in the case of infinite chains), in accordance to Eq. \ref{ldos-s} and
Eq. \ref{ldos-b} respectively. Also, the details on the analytic calculation
of the rates (SC FGR column, obtained by means of the GF poles) are presented
in the Appendix \ref{Apendice-2}.

Systems $\mathbf{I}$ and $\mathbf{II}$ involves only one site. The difference
between them is that $\mathbf{I}$ has a semi infinite chain acting as
environment and $\mathbf{II}$ has an infinite one. The decay rates for these
cases can be directly evaluated within the WBA, and agree exactly with the
numerical solutions.

From the cases $\mathbf{III}$ to $\mathbf{VI}$, the system acquires its own
time scale ($\hbar/V_{AB}$). For these cases, we consider two time scales for
the environment, as compared to the system's time scale. The first one, in
which the time scales are equal ($V=V_{AB}$) and the second in which the
environment is ``accelerated" by $V=5V_{AB}$.

\begin{figure}
    \centering
      \includegraphics[width=0.8\textwidth]{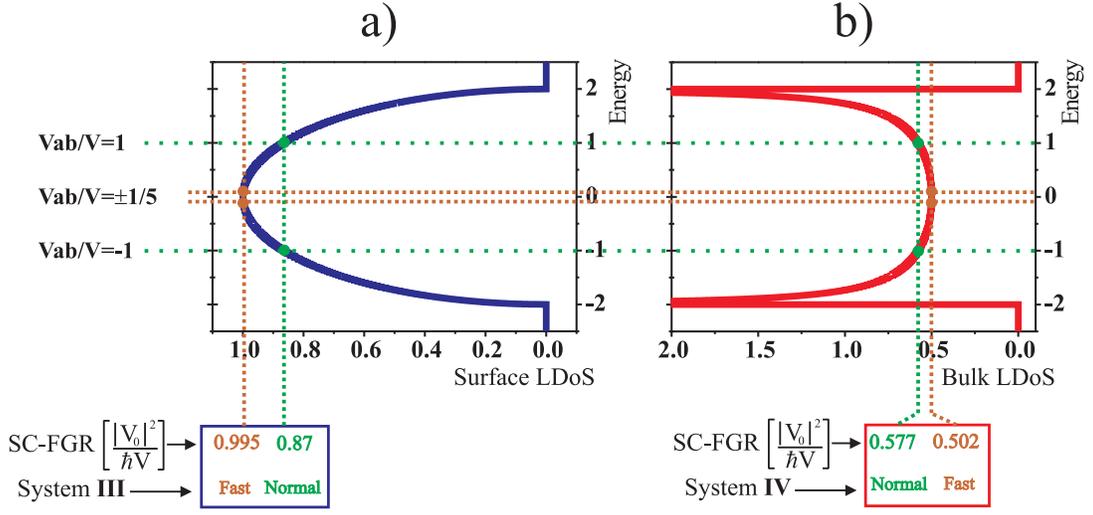}\\
      \caption{(Color Online) Local Density of
States as function of energy (in units of $1/V_{AB}$ and $V_{AB}$ respectively) for: \textbf{a}) Semi-Infinite chain (Surface) and \textbf{b})
Infinite chain (Bulk). The horizontal dotted lines are the values $V_{AB}/V$
which stand for the system energies (renormalized with $V$). The vertical
dotted lines together with their numerical values are the LDoS for each
system in the normal ($V=V_{AB}$) or the fast ($V=5V_{AB}$) configuration.}%
\label{Fig-LDoS}%
\end{figure}

Quite obviously, when the environment has the same time scale as the system,
the WBA rate does not works at all. However, the rates computed by SP and LE
degradation agree with the SC-FGR perfectly (except for the public
environment, case $\mathbf{VI}$, which will be analyzed below). This last
quantity can be interpreted as the LDoS of the bath being evaluated in the
exact solution for the eigen-energies of the whole ($\mathcal{S}$ and
$\mathcal{E}$), and not in the middle of the band spectrum (see Fig.
\ref{Fig-LDoS}-\textbf{a)}. Indeed, the rate of the SC-FGR is obtained by
evaluating the LDoS at the real part of the GF poles.%

It is important to notice that the rate obtained for the SC-FGR are
\textit{higher} or \textit{lower }than the predicted by the WBA, depending if
the bath is \textit{infinite} or \textit{semi-infinite}. We can interpret this
result by looking at Fig. \ref{Fig-LDoS}. Once we evaluate the LDoS in
$\operatorname{Re}\left(  \varepsilon_{pole}\right)  \simeq\pm V_{AB}/V$, we
observe that the values are lower for the SC-FGR than the WBA in the
semi-infinite LDoS, otherwise higher if the infinite LDOS is considered.

At this point, for the two level system, we are able to link the convexity of
the bath's LDoS to the observed decay rate. We notice that if the LDoS is
\textit{convex}, the exact rates are \textit{greater} than the WBA prediction,
but instead if it is \textit{concave,} the rates are \textit{smaller}. In
other words, it depends on the shift away from the middle of the band
spectrum, moving along the LDoS slope.

In order to point to the strictly Markovian case, where any return from
$\mathcal{E}$ to $\mathcal{S}$ is suppressed, we examine accelerated
environments. For these cases, all the rates converge to the same value. From
Fig. \ref{Fig-LDoS}-\textbf{a} we can see the convergence of the exact
solutions towards the middle of the band spectrum, as long as the condition
$V\gg V_{AB}$ is better fulfilled.

So far we have only considered a single and private environment, and in the
following cases we address the private-public discussion. As we mentioned
before, private baths act over each site individually, and a single public
bath acts over two (or eventually more sites) of the same system. This public
$\mathcal{S-E}$ interaction induce new types of correlations, increasing the
dynamical complexity of the physical process.

For case $\mathbf{V}$ we explore the possibility of two private baths (i.e.
two infinite environments each one connected to one site). Since here both
sites, instead of only one, are affected by the environment, it is expected a
double rate compared to the case $\mathbf{IV}$. In Table \ref{tabla-1} we
found that the simulation results agree with the analytical prediction and
again, the rates are greater than the WBA value, which means that we are
moving with the exact eigen-energies through a convex LDoS, away from middle
of the band.

Finally we analyze a very interesting case, where the system is in presence of
a public bath ($\mathbf{VI}$). For these cases it is observed that the SP rate
differs from the local LE. This behavior shows certain asymmetry between the
forward and backward evolutions. Using the SC-FGR approach it can be shown
that the imaginary part of the poles depends on the relative value of $V_{AB}$
and $V$ (see Appendix \ref{Apendice-2}),%

\begin{equation}
\frac{1}{\tau}=\frac{2}{\hbar}\Gamma_{0}\simeq\frac{\sqrt{4V^{2}-V_{AB}^{2}}%
}{2V-V_{AB}}\frac{V_{0}^{2}}{\hbar V}.\label{eq16}%
\end{equation}
This dependence induces a different decay rate whenever the system evolves
forward ($V_{AB}>0$) or backward ($V_{AB}<0$). Thus the SP and the local LE
are not equivalent (see Table \ref{tabla-1}). It is worth mentioning that the
asymmetry (dependence on the relative sign) in the rates for the forward and
backward evolutions, arises only when the bath is public.

The analytical prediction for the local LE rate is obtained observing that the
total evolution for reversed dynamics is proportional to a product of two
exponential evolutions (forward and backward),%

\begin{align}
M_{LE}(T  & =2t)\propto\exp(-\frac{t}{\tau_{f}})\exp(-\frac{t}{\tau_{b}%
})\label{Eq - Product Evol}\\
& =\exp\left[  -t\left(  \frac{\tau_{f}+\tau_{b}}{\tau_{f}\tau_{b}}\right)
\right]  =\exp\left[  -T\left(  \frac{\tau_{f}+\tau_{b}}{2\tau_{f}\tau_{b}%
}\right)  \right]  ,
\end{align}
where $1/\tau_{f}$ and $1/\tau_{b}$ correspond to the forward and backward
degradation rates (see also case $\mathbf{VI}$ in Table \ref{tabla-1}). Eq.
\ref{Eq - Product Evol} shows that the mean rate for a single time-reversed
evolution should agree with the LE decay ($1/\tau_{LE}=1.20\pm0.04$ $V_{0}%
^{2}/\hbar V$). In fact, this is true, since the two SC-FGR rates ($1/\tau
_{f}$ and $1/\tau_{b}$) yield a mean value $1/\tau=1.15$ $V_{0}^{2}/\hbar V$.
Moreover, and quite obviously, the \textit{forward} rate agrees with the SP decay.

\begin{figure}
    \centering
      \includegraphics[width=0.8\textwidth]{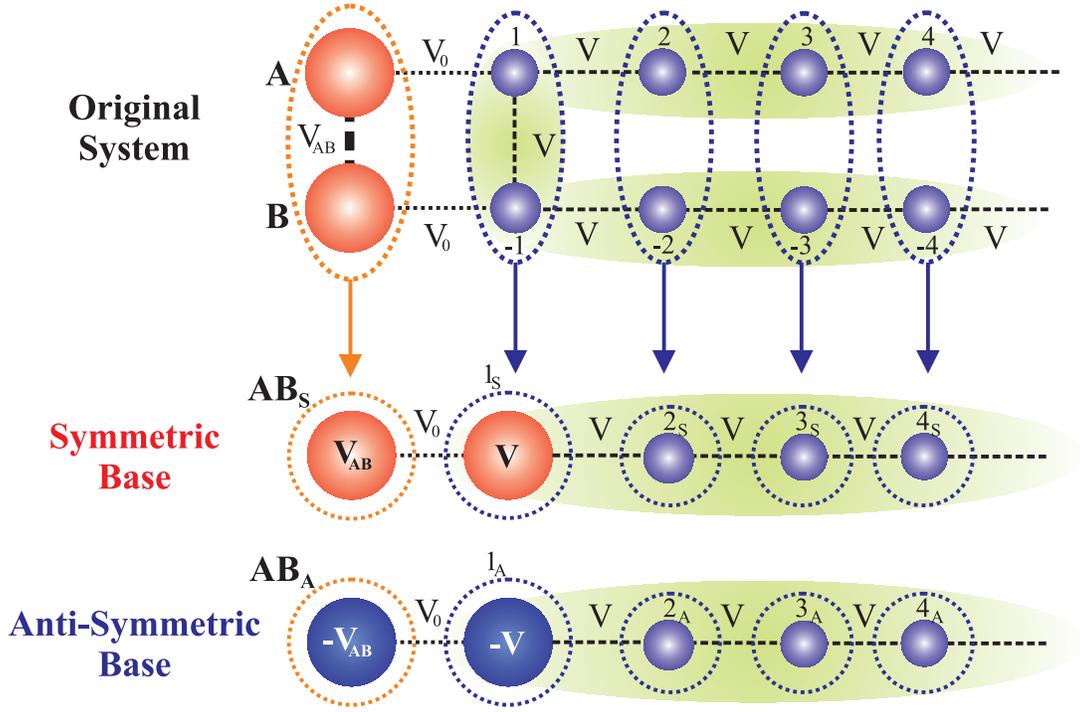}\\
      \caption{(Color online). The schematic
symmetrization procedure of the 2-sites system interacting with the public
environment. The original problem can be cast as two independent semi
infinite linear chains.}%
\label{Fig:Simmetrization}%
\end{figure}

Let us now link the decay process to an appropriate LDoS. As it is clearly
shown in Appendix \ref{Apendice-2}, we can transform the original model of
case $\mathbf{VI}$ to an equivalent one (see Fig. \ref{Fig:Simmetrization}).
If a suitable change of basis is applied (basically turning into symmetric and
anti symmetric basis), then it is only necessary to analyze two semi infinite
linear chains (a particular case treated in Ref. \cite{elenacpl,elenaphysB}).
Since the initial condition has equal weight on both effective chains, the
corresponding rates for them have to be added. Further details on the
symmetrization transformation are explained in Appendix \ref{Apendice-2}.%

\begin{figure}
    \centering
      \includegraphics[width=0.8\textwidth]{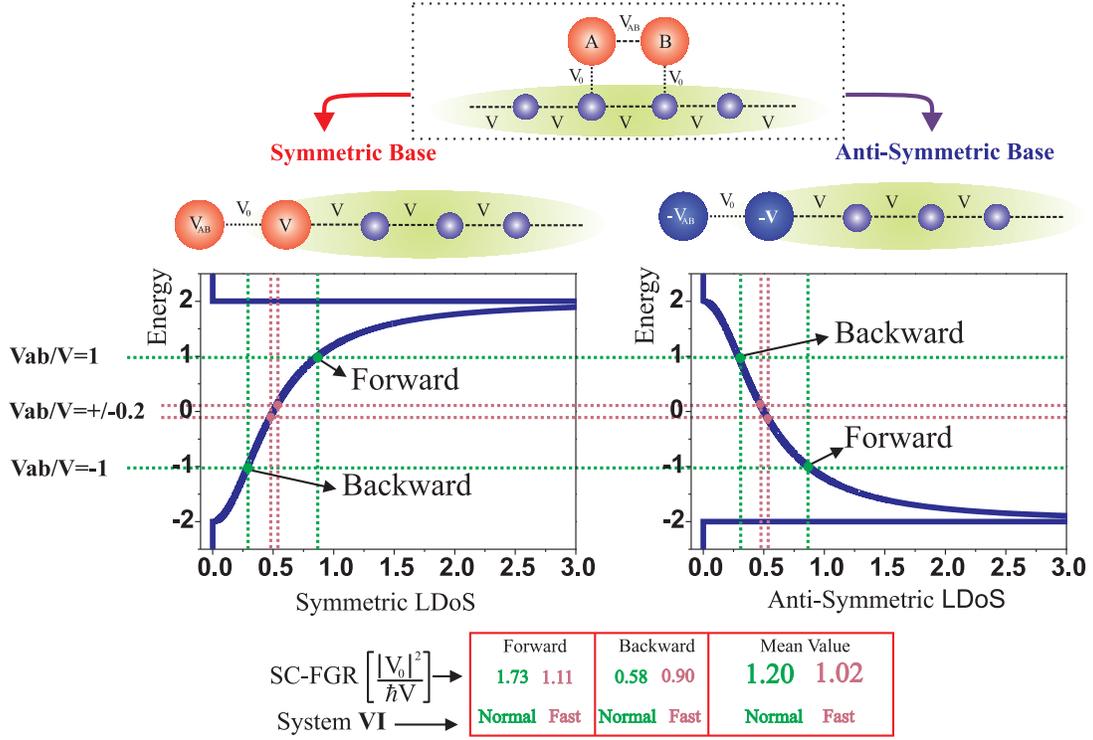}\\
      \caption{(Color online) LDoS as function of energy (in units of $1/V_{AB}$ and $V_{AB}$ respectively) for the
equivalent problem of the public case, after symmetrization transformation.
Forward and backward stages of the evolution are indicated. The horizontal
dotted lines are the values $V_{AB}/V$ which stand for the system
energies (renormalized to $V$). The vertical dotted lines together with
their numerical values are the LDoS for the system \textbf{VI} in the normal
($V=V_{AB}$) or the fast ($V=5V_{AB}$) configuration.}%
\label{public_ldos}%
\end{figure}

From the LDoS considered in Fig. \ref{public_ldos} we can explain why the rate
$1/\tau_{f}$ \ is always above the WBA limit, and the $1/\tau_{b}$ rate is
always below it. Moreover, as long as the \textit{forward} rate agrees with
the SP decay, this also supports the SP rate being greater than the WBA value.

Once more, we make the bath's dynamics faster, leading to a better
applicability of the WBA and the rates converge again to the traditional FGR
description (middle of the band spectrum).%

\begin{figure}
    \centering
      \includegraphics[width=0.8\textwidth]{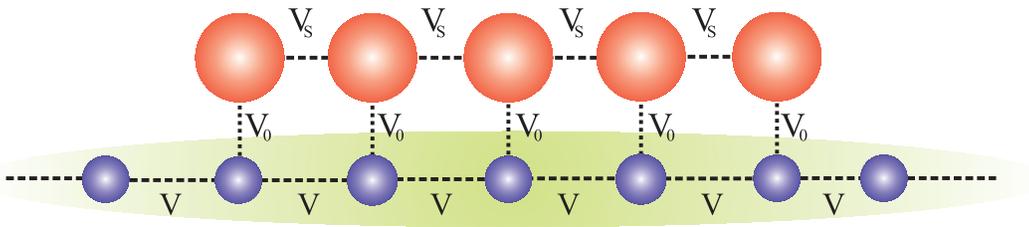}\\
      \caption{(Color Online). Fermion model.
A five-site system laterally coupled to an infinite chain.}%
\label{Fig 5 Sitios}%
\end{figure}

Now we briefly analyze the characteristic local LE decay times for a
\textquotedblleft highly public" bath (See Fig. \ref{Fig 5 Sitios}). This case
is of interest for studying spin dynamics in ring and ladder like systems
(see Ref. \cite{alv2010}).%

In this system we will consider three different relation for time scales:
$V=V_{s}$, $V=5V_{s}$ and $V=8.75V_{s}$ (we also assume $V\gg V_{0}$). For
these cases the decay rates (local LE degradation rate measured in units of
$V_{0}^{2}/\hbar V$) computed by the numerical solution are $(2.66\pm0.07)$,
$(1.54\pm0.05)$ and $(1.16\pm0.04)$, respectively. Therefore, we observe that
the first two results do not match with that expected by WBA. Moreover, as we
have seen before, the convexity of the bulk LDoS, induce bigger values. By
setting $V=8.75V_{s}$ (approximately one order of magnitude of difference) we
are near the WBA validity and thus nearly reproduce the expected result: an
energy independent rate, of value $1.00$ $V_{0}^{2}/\hbar V$, in accordance to
Eq. \ref{s10}. We stress here that the strong sensibility on the relation
between time scales, is enhanced by a highly public $\mathcal{S-E}$ interaction.

\section{FURTHER DISCUSSIONS AND CONCLUSIONS}

\label{sec:Discussions-Conclusions}

Traditionally, the first approach to evaluate decay rates is the FGR based on
an underlying WBA. We have shown several variations of a simple SWAP gate
interacting with an environment (or several ones) where this scheme is not as good as one might
assume. This failure, which can be understood as non-Markovianity, is due to two contributions (which should not be considered as well separated effects).%

First, if the WBA limit ($V\gg V_{AB}$) is not well fulfilled, then it is
quite obvious that the energy independent rate will not be representative as
it would oversimplify the decay process. When the system's time scale becomes
comparable to that of the environment (i.e. $V_{AB}\simeq V$, non-Markovian
situation), the decay rate departs from the usual FGR approximation evaluated
at single energy level. Depending on the bath's spectral structure (LDoS) the
actual rate can be greater or lower. We have shown that if the LDoS is a
convex function of the energy, then any shift away from the middle of the band
spectrum will produce a greater decay rate. But if the LDoS is a concave
energy function, the rate would be lower. In general, we have seen that
dynamical complexity (understood as how much the decay departs from the WBA
validity) arises when the time scales of the system, the environment and the
$\mathcal{S-E}$ interactions, are commensurable.%

Second, when the dynamics is non-Markovian due to similar time scales, the possibility of interaction
through the environment becomes appreciable. In a public bath, the
correlations generated by the multiple connections between different parts of
the system and the bath, are more effective to depart the physical process
away from the WBA validity, for a given relation of time scales. In general
this leads to too subtle correlations to be treated in spectral models for the
bath which here becomes quite natural through the use of specific Hamiltonian
models of the environment.%

In order to address the question on the private and public nature of the $\mathcal{S-E}$
interaction, we confront the cases of \textit{two independent baths} (case $\mathbf{V}$)
and \textit{one common bath} (case $\mathbf{VI}$).
In cases where $\mathcal{S-E}$ interaction is private there is no dependence
of the rate on the relative sign of $V$ and $V_{AB}$. On the other hand, if
the $\mathcal{S-E}$ interaction is public, the forward and backward evolution
will change the decay rates and will produce a mean value for the local LE
decoherence rate. It is important to stress here that we are comparing cases where
$\mathcal{S}$ and $\mathcal{E}$\ have similar time scales. In this condition, it
is observed that decay rates for the public case are greater than for the private.
In this sense we interpret (at least for this cases) that the private bath is less harmful than
a public one. Strictly speaking, this is a consequence of moving along the LDoS
in the \textit{effective} FGR (see Figs. \ref{Fig-LDoS} and \ref{public_ldos}%
). A public scenario produces an enhancement in the alteration of the FGR away
from the WBA limit, as compared to the private one. As a matter of fact,
private bath is less efficient for correlated memory-like returns to the
system. As it has been previously stated in a general sense, and based on
complete formal grounds, the feedback of information from $\mathcal{E}$ to
$\mathcal{S}$ is the central issue for quantifying a non-Markovian bath \cite{breuer}.%

On the other hand, we have shown how Markovianity is restored by changing the relation
between $\mathcal{S}$ and $\mathcal{E}$ time scales. In fact, for all the cases treated in this work,
we have found that once we move towards the WBA limit (represented by a bath dynamics 5 times faster than the
system's), the effects of a public or private bath are no longer relevant.
For both cases the environment is fast enough to wash out all the memory
effects, since any inner excitation is rapidly spread. At this point the rate for both situations become equal and agree
with the one predicted by the common FGR. This means that when the system's
and the bath's time scales are well differentiated (specifically when the
bath's is very short as compared to the system's), then the public-private
reservoir discussion becomes irrelevant.%

In many physical situations, when addressing 1-D and 1-D$^{+}$\ systems (see,
for example \cite{alv2010}), a coupled environment behaves with a convex LDoS.
Therefore, memory effects and complexity in the structure of the
$\mathcal{S-E}$ interaction produce an enhancement of the coherent dynamics
degradation. Additionally, we stress here the wider "spectral exploration" of
the Loschmidt Echo as compared to the Survival Probability (a difference
clearly shown in the public $\mathcal{S-E}$ interaction case, where the global
LE rate is a mean value of two non symmetric processes).%

Even though we do not claim full generality for the results discussed here,
there are some interesting universal issues to remark.
The use of the Wide Band Approximation (simple FGR) can easily lead to quantitative and qualitative errors if the time scales and the physical structure of
$\mathcal{S}$ and $\mathcal{E}$ are similar. Furthermore, the way they are coupled to
each other (\textit{public} or \textit{private}), plays a fundamental role.
In general, dynamical complexity grows when those characteristic times are
similar and when there is no privacy in the $\mathcal{S-E}$ interaction.%

\begin{acknowledgements}
We acknowledge financial support from ANPCyT, SeCyT-UNC, MinCyT-Cor, CONICET.
This work was benefited from fruitful discussions with P. R. Levstein and R. Bustos-Mar\'{u}n.
The authors acknowledge F. Rojo and F. Pastawski for useful comments on the manuscript.
\end{acknowledgements}

\appendix
\section{Excitation dynamics in 1-d systems and the Spin-Fermion mapping}

\label{Apendice-1} In previous sections we developed a scheme that could be
used to treat spin polarization dynamics \cite{madi-ernst1997,mesoECO-theory},
under certain assumptions. In that sense, the well-known mapping between spins
and fermions \cite{jwt} has been used for formulating the spin problem in
terms of the non-equilibrium Keldysh formalism
\cite{keldysh,danieli1,Danieli-SWAP-decoh}. We briefly present here the Spin-Fermion
mapping, and how the tight binding models discussed along the present article arise.

A simple case that can be treated in this context is a linear chain of $M$
spins in an external magnetic field. They interact with their nearest
neighbors through $XY$ coupling:%

\begin{align}
\hat{H} &  =\sum_{n=0}^{M-1}\hbar\Omega_{n}\hat{S}_{n}^{z}-\sum_{n=0}%
^{M-2}J_{n+1,n}[\hat{S}_{n+1}^{x}\hat{S}_{n}^{x}+\hat{S}_{n+1}^{y}\hat{S}%
_{n}^{y}]\label{s3}\\
&  =\sum_{n=0}^{M-1}\hbar\Omega_{n}\hat{S}_{n}^{z}-\sum_{n=0}^{M-2}\frac{1}%
{2}J_{n+1,n}[\hat{S}_{n+1}^{+}\hat{S}_{n}^{-}+\hat{S}_{n+1}^{-}\hat{S}_{n}%
^{+}],\label{s4}%
\end{align}
\qquad

where $\hat{S}_{n}^{\pm}$ are the rising and lowering operators $\hat{S}%
_{n}^{\pm}=\hat{S}_{n}^{x}\pm \mathrm{i} \hat{S}_{n}^{y}$. The dynamics of the $M$-spin
system, evolving under the Hamiltonian $\hat{H}$, is usually described by
means of the two site spin correlation function,%

\begin{equation}
P_{f,i}(t)=\frac{\left\langle \Psi_{eq}\right\vert \hat{S}_{f}^{z}(t)\hat
{S}_{i}^{z}(t_{0})\left\vert \Psi_{eq}\right\rangle }{\left\langle \Psi
_{eq}\right\vert \hat{S}_{f}^{z}(t_{0})\hat{S}_{i}^{z}(t_{0})\left\vert
\Psi_{eq}\right\rangle }.\label{s1}%
\end{equation}
The quantity of Eq. \ref{s1} gives the amount of local polarization in the $z
$ component at time $t$ on the $f$th site, provided that the system was, at
time $t_{0}$, in its equilibrium state with a spin \textit{up} added at $i $th
site. Also, $\hat{S}_{f}^{z}(t)=e^{\mathrm{i}\hat{H}t}\hat{S}_{f}^{z}e^{-\mathrm{i}\hat{H}t}$ is
the spin operator in the Heisenberg representation and $\left\vert \Psi
_{eq}\right\rangle =\sum_{N}a_{N}\left\vert \Psi_{eq}^{(N)}\right\rangle $ is
the thermodynamical many-body equilibrium state constructed by adding states
with different number $N$ of spins \textit{up }with appropriate statistical
weights and random phases.%

The Jordan-Wigner transformation (JWT) links spin and fermion operators at
each site, by the following relation:%

\begin{equation}
\hat{S}_{n}^{+}=\hat{c}_{n}^{\dag}\exp \left[   \mathrm{i}\pi\sum_{m=1}^{n-1}\hat{c}%
_{m}^{\dag}\hat{c}_{m}^{{}} \right]  \label{s2}%
\end{equation}
where $\hat{c}_{m}^{\dag}$, $\hat{c}_{m}$ are the canonical fermionic
operators. The use of the JWT on the Hamiltonian \ref{s4} yields:%

\begin{equation}
\hat{H}=\sum_{n=0}^{M-1}\varepsilon_{n}(\hat{c}_{n}^{\dag}\hat{c}_{n}-\frac
{1}{2})-\sum_{n=0}^{M-2}V_{n+1,n}[\hat{c}_{n+1}^{\dag}\hat{c}_{n}^{{}}+\hat
{c}_{n}^{\dag}\hat{c}_{n+1}^{{}}],\label{s5}%
\end{equation}
where $\varepsilon_{n}\equiv\hbar\Omega_{n}$ are the site energies and
$V_{n+1,n}\equiv\frac{1}{2}J_{n+1,n}$ are the hoppings. Due to the short range
interaction (first neighbors), after the application of the JWT, the only
non-zero coupling terms between spins are proportional to $\hat{c}_{n+1}%
^{\dag}\hat{c}_{n}=\hat{S}_{n+1}^{+}\hat{S}_{n}^{-}$. Each subspace with
$\binom{M}{N}$ states of spin projection $\left\langle \sum_{n=1}^{M}\hat
{S}_{n}^{z}\right\rangle =N-M/2$ is now a subspace with $N$ non-interacting
fermions. The eigenfunctions $\left\vert \Psi_{\gamma}^{(N)}\right\rangle $
are expressed as a single Slater determinant built up with the single particle
wave functions $\varphi_{\alpha}$ of energy $\varepsilon_{\alpha}$. Under
these circumstances, and setting $\left\vert i\right\rangle \equiv\hat{c}%
_{i}^{\dag}\left\vert \emptyset\right\rangle $ (with $\left\vert
\emptyset\right\rangle $ the fermion vacuum), Eq. \ref{s1} reduces to:%

\begin{align}
P_{f,i}(t) &  =\left\vert \left\langle f\right\vert \exp\left[  -\mathrm{i}\hat
{H}t/\hbar\right]  \left\vert i\right\rangle \Theta(t)\right\vert
^{2}\label{s6}\\
&  =\hbar^{2}\left\vert G_{f,i}^{R}(t)\right\vert ^{2},\nonumber
\end{align}
where $G_{f,i}^{R}(t)$ is the retarded Green's function for a single fermion
that connects sites $f$ and $i$.

While similar steps lead to description of excitations in Double Quantum (DQ)
Hamiltonian \cite{feldman1,feldman2,cory2,elenaMQC}:
\begin{equation}
\hat{H}_{DQ}=\sum_{n=0}^{M-2}\frac{J_{n+1,n}}{2}[\hat{S}_{n+1}^{+}\hat{S}%
_{n}^{+}+\hat{S}_{n+1}^{-}\hat{S}_{n}^{-}],\label{hdq}%
\end{equation}
those will not be detailed here. The fundamental issue is the underlying
one-body dynamics, which for the DQ Hamiltonian is revealed by a unitary
transformation $\hat{H}_{DQ}=U^{\dag}\hat{H}_{XY}U$ \cite{feldman2,elenaMQC}
that links it to an $XY$ \ Hamiltonian.

Hence, for one dimensional (1-D) chains of spins with first neighbors $XY$ or
DQ interactions, and in the high temperature regime, the dynamics of an
excitation (either an injected local polarization or multiple quantum
coherence) is completely equivalent to the evolution of a single particle wave
function, ruled by a tight-binding Hamiltonian. Therefore, it turns out that
the analysis of Hamiltonians like the one in Eq. \ref{s5}, can be casted for
treating and studying several effects in spin chains \cite{viola} (a typical
scenario in QIP).

\section{Green's Function Poles.}

\label{Apendice-2} Here we present the detailed analytical derivation of the
GF poles and further approximations. The results from this appendix have been
summarized in the Table \ref{tabla-1} showed in Sec. \ref{sec:Results}.

The first model under consideration (Fig. \ref{Fig Esquema de Sistemas}%
-\textbf{I}) is given by one site coupled to a semi-infinite chain. In this
case the GF pole results purely imaginary,%

\begin{equation}
\varepsilon_{pole}=-\mathrm{i}\frac{V_{0}^{2}}{\sqrt{V^{2}-V_{0}^{2}}}%
\text{.}\label{a1}%
\end{equation}
Even though this is indeed the exact solution, we span it for $V_{0}^{2}%
/V\ll1$ and obtain:%

\begin{equation}
\varepsilon_{pole}\simeq\frac{V_{0}^{2}}{V}\text{ .}\label{a2}%
\end{equation}
Thus the theoretical decay rate for this model can be expressed as:%

\begin{equation}
\frac{1}{\tau}=\frac{2}{\hbar}\Gamma_{0}=\frac{2}{\hbar}\operatorname{Im}%
(\varepsilon_{pole})\simeq\frac{2}{\hbar}\frac{V_{0}^{2}}{V}.\label{a3}%
\end{equation}
If we change the environment to an infinite chain (Fig.
\ref{Fig Esquema de Sistemas}-\textbf{II}) the pole is,%

\begin{equation}
\varepsilon_{pole}=-\mathrm{i}\left(  2V^{2}(1-\sqrt{\frac{V_{0}^{4}}{4V^{4}%
}+1})\right)  ^{1/2}\simeq-\mathrm{i}\frac{V_{0}^{2}}{2V}\label{a4}%
\end{equation}
The case \textbf{III,} which corresponds to two sites coupled to a
semi-infinite chain, has been solved previously in Ref. \cite{axel},%

\begin{equation}
\varepsilon_{pole}^{2}=\frac{V_{AB}^{2}\left(  2V^{2}-V_{0}^{2}\right)
-V_{0}^{4}\pm V_{0}^{2}\sqrt{\left(  V_{AB}^{2}+V_{0}^{2}\right)  ^{2}%
-4V_{AB}^{2}V^{2}}}{2\left(  V^{2}-V_{0}^{2}\right)  }.\label{a5}%
\end{equation}
Using the solutions of Eq. (26) and (27) in Ref. \cite{axel}, we can directly
evaluate the real and imaginary parts of the poles,%

\begin{align}
\Delta_{0}  &  =\pm\left(  \frac{V_{AB}^{2}\left(  2V^{2}-V_{0}^{2}\right)
-V_{0}^{4}}{2\left(  V^{2}-V_{0}^{2}\right)  }+\Gamma_{0}^{2}\right)
^{1/2}\nonumber\\
&  \simeq\pm V_{AB}\left(  1-\frac{1}{4V}\frac{V_{0}^{2}}{V}\right)
\label{a6}\\
\Gamma_{0}  &  =\left(  \frac{V_{0}^{4}-V_{AB}^{2}(2V^{2}-V_{0}^{2})}%
{4(V^{2}-V_{0}^{2})}+\sqrt{\frac{V^{2}V_{AB}^{4}}{4(V^{2}-V_{0}^{2})}}\right)
^{1/2}\nonumber\\
&  \simeq-\frac{1}{4}\frac{\sqrt{4V^{2}-V_{AB}^{2}}}{V}\frac{V_{0}^{2}}%
{V}.\label{a7}%
\end{align}
Now, we change slightly the geometry of these systems and consider again an
infinite chain as environment instead the semi-infinite. For these cases we
will write only the first non trivial terms of their Taylor expansion instead
of the full solution. For the model of Fig. \ref{Fig Esquema de Sistemas}%
-\textbf{IV,} the solution is:%

\begin{align}
\Delta_{0}  &  \simeq\pm V_{AB}+O(\frac{V_{0}^{4}}{V^{4}})\label{a8}\\
\Gamma_{0}  &  \simeq-\frac{1}{2}\frac{V}{\sqrt{4V^{2}-V_{AB}^{2}}}\frac
{V_{0}^{2}}{V}+O(\frac{V_{0}^{4}}{V^{4}}).\label{a9}%
\end{align}

The next case (Fig. \ref{Fig Esquema de Sistemas}-\textbf{V)} is a two-site
system coupled to two private baths. In this case, the solution is,%

\begin{align}
\Delta_{0}  &  \simeq\pm V_{AB}+O(\frac{V_{0}^{4}}{V^{4}})\label{a10}\\
\Gamma_{0}  &  \simeq-\frac{V}{\sqrt{4V^{2}-V_{AB}^{2}}}\frac{V_{0}^{2}}%
{V}+O(\frac{V_{0}^{4}}{V^{4}}).\label{a11}%
\end{align}
At this point we observe that the imaginary part for the case \textbf{V} is
twice of the system \textbf{IV}. This behavior is consistent with the count of
the ``number" of private baths connected to the system, and the proportion
affected by those private baths.

Finally, for the case of Fig. \ref{Fig Esquema de Sistemas}-\textbf{VI}, the
model involves a two-site system coupled to a common bath. For this case the
solution is expressed in the following form,%
\begin{align}
\Delta_{0}  &  \simeq\pm\left(  V_{AB}-\frac{1}{2}\frac{V_{0}^{2}}{V}\right)
+O(\frac{V_{0}^{4}}{V^{4}})\label{a13}\\
\Gamma_{0}  &  \simeq-\frac{\sqrt{4V^{2}-V_{AB}^{2}}}{4V-2V_{AB}}\frac
{V_{0}^{2}}{V}+O(\frac{V_{0}^{4}}{V^{4}}).\label{a14}%
\end{align}%

Notice that the imaginary part of the pole is linear in $V_{AB}$ in a sign dependent manner (in contrast to previous cases). This linearity
translate into a different value for $\Gamma_{0}$ depending on the sign of
$V_{AB}$. As a matter of fact, the local LE indeed relies on the change of
sign to revert the dynamics. To understand the physics below this difference
we have to identify the LDoS involved in the decay process.
Accordingly, we symmetrize the basis, thus dimmerizing the system, as shown in Fig. \ref{Fig:Simmetrization} (i.e. we take pairs of site states and map them into symmetric and anti symmetric states.)
Therefore the tight binding Hamiltonian for this case is mapped to the form of Eq.
\ref{hamilt_simetriz}.%

\begin{align}
\hat{H}  & =
\bordermatrix{~ & \left\vert A\right\rangle & \left\vert B\right\rangle & \left\vert 1\right\rangle & \left\vert -1\right\rangle & \left\vert 2\right\rangle & \left\vert -2\right\rangle & \cdots & \cdots \cr \left\langle A\right\vert & & V_{AB} & & V_{0} & & & & \cr \left\langle B\right\vert & V_{AB} & & V_{0} & & & & & \cr \left\langle 1\right\vert & & V_{0} & & V & V & & & \cr \left\langle -1\right\vert & V_{0} & & V & & & V & & \cr \left\langle 2\right\vert & & & V & & & & \ddots & \cr \left\langle -2\right\vert & & & & V & & & & \ddots\cr \vdots & & & & & \ddots & & & \cr \vdots & & & & & & \ddots & & \cr}\nonumber\\
\longrightarrow\hat{H}^{\prime}  & =
\bordermatrix{~ & \cdots & \left\vert 2_{S}\right\rangle & \left\vert 1_{S}\right\rangle & \left\vert AB_{S}\right\rangle & \left\vert AB_{A}\right\rangle & \left\vert 1_{A}\right\rangle & \left\vert 2_{A}\right\rangle & \cdots\cr \vdots & & \ddots & & & & & & \cr \left\langle 2_{S}\right\vert & \ddots & & V & & & & & \cr \left\langle 1_{S}\right\vert & & V & V & V_{0} & & & & \cr \left\langle AB_{S}\right\vert & & & V_{0} & V_{AB} & & & & \cr \left\langle AB_{A}\right\vert & & & & & -V_{AB} & V_{0} & & \cr \left\langle 1_{A}\right\vert & & & & & V_{0} & -V & V & \cr \left\langle 2_{A}\right\vert & & & & & & V & & \ddots\cr \vdots & & & & & & & \ddots & \cr}\label{hamilt_simetriz}%
\end{align}

where $\left\vert n_{S}\right\rangle =$ $\left(  \left\vert n\right\rangle
+\left\vert -n\right\rangle \right)  /\sqrt{2}$, $\left\vert n_{A}%
\right\rangle =$ $\left(  \left\vert n\right\rangle -\left\vert
-n\right\rangle \right)  /\sqrt{2},$\ $n=1,2,3..$and $\left\vert
AB_{S}\right\rangle =$ $\left(  \left\vert A\right\rangle +\left\vert
B\right\rangle \right)  /\sqrt{2}$, $\left\vert AB_{A}\right\rangle =\left(
\left\vert A\right\rangle -\left\vert B\right\rangle \right)  /\sqrt{2}$. From
the Hamiltonian $\hat{H}^{\prime}$ (Eq. \ref{hamilt_simetriz}) it is easy to
identify the splitting of the original problem into two semi infinite tight
binding chains, with only the first two site energies non zero. This problem
has been previously addressed in Refs. \cite{elenacpl} and \cite{elenaphysB}
and we recall the LDoS computed there. Thus in Fig. \ref{public_ldos} we
identify the energy excitation on the spectral structure of the environment,
which is relevant during the decay process.

%\begin{thebibliography}{99}
%\bibliography{CitasBibi}

\begin{thebibliography}{75}
\bibitem [{com()}]{comment1}%
  \BibitemOpen
  \href@noop {} {\ }\bibinfo {note} {Note: Notice that the systems we are
  considering do not have enough complexity to warrant a perturbation
  independent (i.e. Lyapunov) decoherence rate.}\BibitemShut {Stop}%
\end{thebibliography}
%\end{thebibliography}

%

\end{document}